\documentclass[a4paper]{article}
\usepackage[hidelinks]{hyperref}
\usepackage{INTERSPEECH2022}

\title{KaraTuner: Towards End-to-End Natural Pitch Correction for Singing Voice in Karaoke}
\name{Xiaobin Zhuang$^{1}$, Huiran Yu$^{2}$\thanks{Huiran Yu performed the work during her internship at Tencent Music Entertainment Lyra Lab. Xiaobin Zhuang and Huiran Yu made an equal contribution to the article.}, Weifeng Zhao$^{1}$, Tao Jiang$^{1}$, Peng Hu$^{1}$}
\address{
  $^{1}$ Tencent Music Entertainment Lyra Lab, Shenzhen, China \\
  $^{2}$ Carnegie Mellon University, Pittsburgh, PA, USA}
\email{$^{1}$\{aaronzhuang, ethanzhao, marsjiang, stevenhu\}@tencent.com \\ $^{2}$huiranyu@andrew.cmu.edu}

\begin{document}
\maketitle
%%%%%%%%%%%%%%%%%%%%%%%%%%%%%%%%%%%%%%%%%%%%%%%%%%%%%%%%%%%%%
\begin{abstract}
An automatic pitch correction system typically includes several stages, such as pitch extraction, deviation estimation, pitch shift processing, and cross-fade smoothing. However, designing these components with strategies often requires domain expertise and they are likely to fail on corner cases. In this paper, we present KaraTuner, an end-to-end neural architecture that predicts pitch curve and resynthesizes the singing voice directly from the tuned pitch and vocal spectrum extracted from the original recordings. Several vital technical points have been introduced in KaraTuner to ensure pitch accuracy, pitch naturalness, timbre consistency, and sound quality. A feed-forward Transformer is employed in the pitch predictor to capture long-term dependencies in the vocal spectrum and musical note.
We also develop a pitch-controllable vocoder based on a novel source-filter block and the Fre-GAN architecture.  
KaraTuner obtains a higher preference than the rule-based pitch correction approach through A/B tests, and perceptual experiments show that the proposed vocoder achieves significant advantages in timbre consistency and sound quality compared with the parametric WORLD vocoder, phase vocoder and CLPC vocoder.
\end{abstract}

%%%%%%%%%%%%%%%%%%%%%%%%%%%%%%%%%%%%%%%%%%%%%%%%%%%%%%%%%%%%%
\begin{figure*}[ht]
  \centerline{\includegraphics[width=14cm]{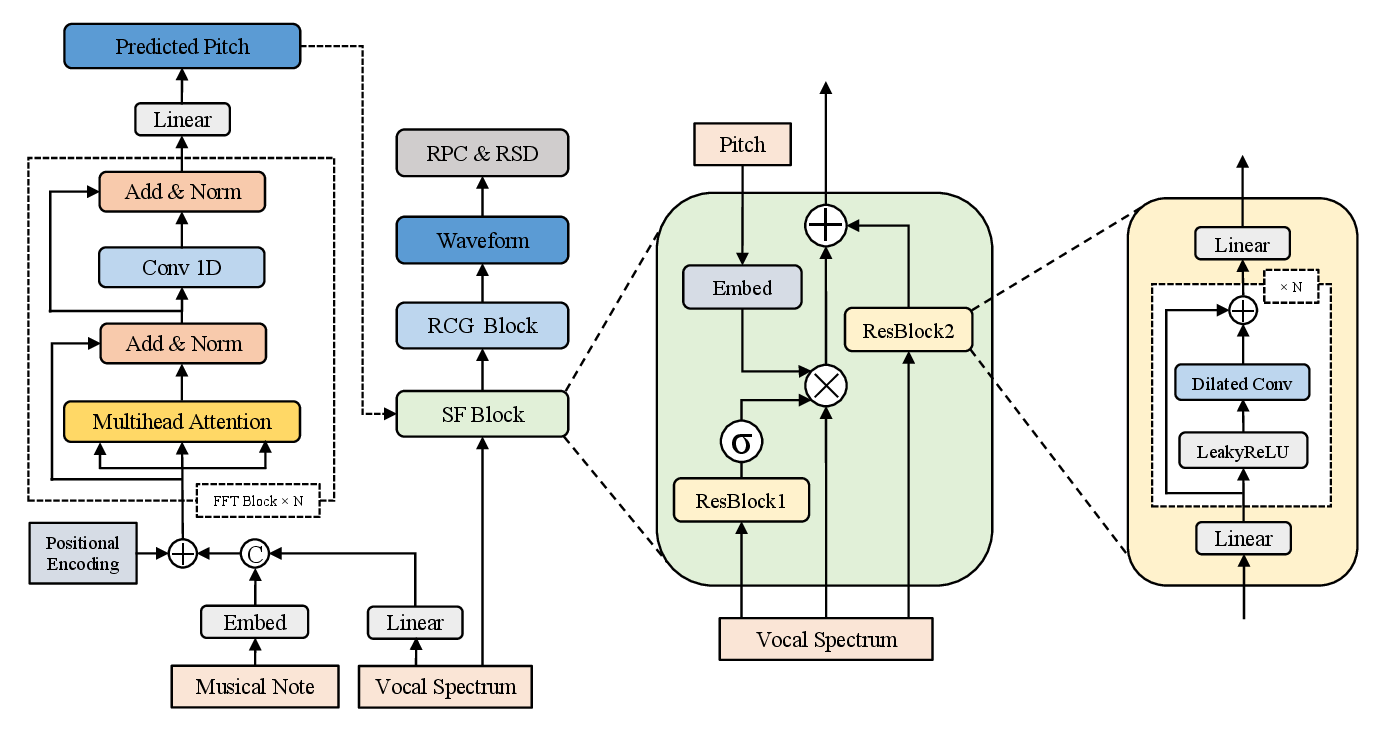}}
  \caption{The overview of the proposed KaraTuner. The \textit{C} operator denotes feature concatenation and linear projection. The \textit{+} operator and the \textit{×} operator denote the feature addition and feature multiplication, respectively. \textit{FFT Block × N} denotes that the FFT Block repeats \textit{N} times. The building block in the ResBlock repeats \textit{N} times, with different dilate factors of convolution. The RCG, RPC, and RSD blocks utilize the structure from the Fre-GAN vocoder with specific hyperparameters.}
  \label{fig:karatuner}
  \end{figure*}
%%%%%%%%%%%%%%%%%%%%%%%%%%%%%%%%%%%%%%%%%%%%%%%%%%%%%%%%%%%%%
\section{Introduction}
%%%%%%%%%%%%%%%%%%%%%%%%%%%%%%%%%%%%%%add(这一段帮忙看看)

Pitch correction is a widely applied voice editing technique, as it improves the intonation of the singers and helps create professional music products. 
In the music production industry, pitch correction is often performed by professional music engineers with sufficient domain knowledge using commercial pitch correction tools such as Melodyne and Autotune. In recent years, there has been a growing interest in developing automatic pitch correction algorithms among researchers.
%%%%%%%%%%%%%%%%%%%%%%%%%%%%%%%%%%%%%%%%%%%

A common idea to improve singing performance is to adopt features from professional singers with the help of time warping algorithms. 
Luo et al. \cite{luo2018singing} proposed a canonical time warping algorithm \cite{Zhou2009CanonicalTW} that combines the canonical correlation analysis with dynamic time warping to port pitch curves from professional recordings into user singing.
Yong et al. \cite{8461793} further transferred energy dynamics from professional singing.
%%%%%%%%add
Recently, Liu et al. \cite{liu2022learning} proposed a novel Shape-Aware DTW (SADTW) algorithm, which ameliorates the robustness of existing time-warping approaches by considering the shape of the pitch curve rather than low-level features when calculating the optimal alignment path. A latent-mapping algorithm was also designed to improve the vocal tone of the voice.
%%%%%%%%%%%
However, deeply relying on a voice reference, in real-world applications these methods suffer from difficulties in template acquisition and their tuned performances are inevitably homogeneous in singing style.
The data-driven approach proposed by Wager et al. \cite{wager2020deep} predicts pitch shifts from the difference between the singing voice and the accompaniment, which keeps the singing style to a greater extent and eases the homogeneity problem. 
However, the pitches identified from the accompaniment may not be accurate enough, and the pitch deviation is difficult to assess when the singer is severely off the correct melody.
Score-based approaches like \cite{perrotin:hal-01672238} and \cite{6855142} usually use a set of rules to generate a target pitch curve from the given MIDI sequence. Although a note template is convenient to produce and is more reliable than the accompaniment, these strategies require careful parameter tuning and are not robust with corner cases.
%%%%%%%%%%%%%%%%%%%%%%%%%%%%%%%%%%%%%%add(这一段帮忙看看)

In addition to relocating the pitch curve, another vital part of pitch correction system is resynthesizing the signal with the new tuned pitch, where a pitch-controllable vocoder is essential. 
Methods based on digital signal processing (DSP) such as phase vocoder \cite{phasevocoder}, SOLA \cite{VERHELST2000207} \cite{esola}, and WORLD\cite{Morise2016WORLDAV} vocoder are feasible for the task. However, they tend to introduce artifacts and robotic voice into the synthesized audio. In recent years, neural network-based audio synthesis methods have received increasing attention. Differentiable DSP (DDSP) \cite{engel2020ddsp} has been introduced as a new method to generate audio with deep learning, where DSP algorithms are used as part of a neural network, ensuring end-to-end optimization. Since the first published examples of DDSP were focused on timbre transfer from monophonic instruments, Alonso et al. \cite{alonso2021latent} present the DDSP architecture to a more complex, expressive instrument: the human vocal apparatus and check the suitability of the DDSP for singing voice synthesis by conditioning the model on the Mel Frequency Cepstral Coefficents (MFCC) of the original audio and creating a latent space. Other neural vocoders include WaveNet \cite{oord2016wavenet}, WaveRNN \cite{kalchbrenner2018efficient}, WaveGlow \cite{prenger2019waveglow} and Parallel WaveGAN \cite{yamamoto2020parallel} do not address pitch-shifting problem, while LPCNet \cite{valin2019lpcnet} which resembles a source-filter model, has the capability of pitch-shifting and exhibits more natural timbre than traditional phase vocoders \cite{quatieri2006discrete}. Based on LPCNet, Morrison et al. \cite{morrison2021neural} proposed Controllable LPCNet (CLPCNet), an improved LPCNet vocoder capable of pitch-shifting and time-stretching of speech.
%%%%%%%%%%%%%%%%%%%%%%%%%%%%%%%%%%%%%%%%%%%%

%A vocoder is used to resynthesize the signal with the new tuned pitch, where pitch controllable synthesis method such as phase vocoder, SOLA and WORLD\cite{Morise2016WORLDAV} vocoder are common choices. However, these vocoders tend to introduce defects and robotic voice into the synthesized audio. %which needs to be separated from the original sound track or additionally recorded. These separated vocal tracks sometimes suffer from frequency band missing and can not be used as the reference. Another approach is to conduct pitch correction without explicit target pitch but only referring to the backing track.

To overcome the drawbacks of the above methods, we propose KaraTuner, a novel architecture for automatic pitch correction in karaoke.
%that predicts a new pitch curve from the vocal spectrum and the reference score, then synthesizes the audio with a neural network vocoder based on Fre-GAN \cite{kim2021fregan}. 
The main contributions of our work are as follows:
1) We propose a vocal-adaptable pitch predictor to replace the rule-based pitch shift strategies to achieve diversity and naturalness of the predicted pitch. %in auto-tuning systems\\.
2) We develop a source-filter (SF) block to achieve pitch controllability. We use the pitch-integrated hidden representation from the SF block instead of the Mel-spectrogram as the activation of the neural vocoder. 
3) We propose a practical data preprocessing method to build dataset from unlabeled amateur singing instead of any professional recordings.
%since the spectral envelope contains many(more?) aspects of information such as lyric content, singer timbre and music interpretation. 
In the experiments, we use the rule-based approach and existing vocoders as the baseline, to show that KaraTuner is superior in pitch accuracy, pitch naturalness, timbre consistency, and sound quality.

%%%%%%%%%%%%%%%%%%%%%%%%%%%%%%%%%%%%%%%%%%%%%%%%%%%%%%%%%%%%%
\section{ARCHITECTURE}
Figure \ref{fig:karatuner} illustrates the architecture of KaraTuner. We set up a pitch predictor with the Feed-Forward Transformer \cite{vaswani2017attention} (FFT) blocks and a pitch-controllable vocoder based on a source-filter block and the Fre-GAN architecture. In the training phase, these two modules are trained separately. Meanwhile, the ground truth pitch rather than the predicted pitch is passed through the source-filter block to maintain pitch consistency between the input and output of the vocoder for faster convergence.
%%%%%%%%%%%%%%%%%%%%%%%%%%%%%%%%%%%%%%%%%%%%%%%%%%%%%%%%%%%%%
\subsection{Vocal-Adaptable Pitch Predictor}
In speech and singing voice synthesis, people usually consider the spectral envelope as the timbre representation of a speaker or singer, and its relationship with the pitch curve is generally ignored. However, En-Najjary et al. \cite{EnNajjary2003ANM} reported that the spectral envelope feature implicitly contains the pitch curve, as they predicted it out of the spectral envelope with high accuracy. Inspired by this work, we take into account the spectral envelope and develop a vocal-adaptable pitch predictor to customize in-tune natural pitch curves.
%The frequencies of the notes in the score 
%The musical note in linear frequency scale first go through an embedding layer, and then are concatenated together with the linear projection of spectral envelop feature. 
The input of the pitch predictor consists of the musical note and vocal spectrum. Here, the vocal spectrum is the spectral envelope feature in uncompressed linear scale for complete information. The note embeddings and the linear projection of the vocal spectrum are concatenated and then fed into a stack of FFT blocks. Finally, a linear projection layer is added to map the dimensions of output hidden features and the target pitch. We do not adopt a residual connection between the input notes and the output pitch\cite{lu2020xiaoicesing}, since experiments show that the residual connection will introduce breakpoints at the transition of notes.
%The network has to learn to generate inverse jumps, which is not feasible. (Fig.) We let the network learn how to conduct a smooth transition instead.
% The encoder’s inputs first flow through a self-attention layer – a layer that helps the encoder look at other words in the input sentence as it encodes a specific word. We’ll look closer at self-attention later in the post.
% The outputs of the self-attention layer are fed to a feed-forward neural network. The exact same feed-forward network is independently applied to each position.
Since the spectral envelope implicitly contains the pitch curve, we randomly shifted the spectral envelope along the frequency axis in the training phase to alleviate over-fitting and force the reference score to be the backbone of the pitch curve and the spectral envelope to express details such as gliding and vibrato. In our pitch prediction task, the information related to the pitch curve is concentrated in the middle-low frequency bands of the spectral envelope. Therefore, we drop the redundant high-frequency features. 
%`The spectral envelop feature only controls the \textit{details} of the pitch curve. This guides the generated curve to move around the target score.`
Finally, we use mean squared error (MSE) loss between the predicted pitch curve $\hat{x}$ and the ground truth $x$ to optimize the pitch predictor. The MSE loss of the pitch predictor is deﬁned as: 
\begin{equation}
    \mathcal{L}_{MSE} = \mathbb{E}[||x - \hat{x}||_2]
\end{equation}
%%%%%%%%%%%%%%%%%%%%%%%%%%%%%%%%%%%%%%%%%%%%%%%%%%%%%%%%%%%%%(这一段帮忙看看)
\subsection{Pitch-Controllable Neural Vocoder}

Most neural network vocoders cannot maintain the f0-consistency of the waveform, and many perform better on single-speaker datasets. At the same time, the sound quality usually downgrades when they generate audio of unseen speakers. Therefore, we adopted the universal neural vocoder Fre-GAN structure for high-fidelity any-speaker waveform generation. To further integrate pitch controllability, we designed a neural source-filter block inspired by WORLD vocoder\cite{world} and \cite{wang2019neural}, based on the assumption that the source is independent from the filter, and human voice can be synthesized by convolving the source signal with the filter impulse response. Besides, SingGAN vocoder by Chen et al. \cite{chen2021singgan} also indicates that the use of pitch condition helps synthesize waveforms with stable and natural vowel pronunciation, which improves the audio quality. Hence, we developed a novel neural source-filter block, which combines the pitch feature with vocal spectrum envelope and also alleviates the glitch problem in the spectrogram.

%%%%%%%%%%%%%%%%%%%%%%%%%%%%%%%%%%%%%%%%%%%%%%%%%%%%%%%%%%%%%
\subsubsection{Source-Filter Block}
\label{ssec:SFblock}
% The acoustics of human speech production is typically represented by a linear
% source-filter (SF) model. This linear model is based on the assumption that the source
% is independent of the filter (Fant, 1971), where the source signal is convolved with the
% filter impulse response to generate speech.
%https://asa.scitation.org/doi/pdf/10.1121/1.5049510
% The WORLD vocoder decomposes the signal into spectral envelope, aperiodicity, and pitch p for every frame comprising Ns samples (Morise et al.,
% 2016). The quality of this decomposition depends on the accuracy of the pitch estimate
% (Morise et al., 2016). Hence we estimate the pitch from the Differentiated Electro
% Glotto Graph signal, based on which the spectral envelope, aperiodicity, are estimated
% using the WORLD vocoder.
%and can be resynthesized by a source-filter model. 
In KareTuner, the inputs of the source-filter (SF) block are the pitch curve and the spectral envelope. In the training phase, the ground truth pitch is directly fed into the SF block, while in the inference phase, the predicted pitch is masked with the voiced/unvoiced (V/UV) decision of the original audio before feeding into the network. 

A vocal signal \textit{s} typically consists of periodic and aperiodic components. In the SF block, the pitch goes through an embedding layer and does element-wise multiplication with the spectral envelope to generate the periodic component. Independently, the spectral envelope also goes through a ResBlock2 to predict the aperiodic component. 
A simple way to combine these two components is to add them directly. However, we found that %the additional use of residual connections to estimate the 
a learnable mixing ratio of each frame can improve the sound quality of synthesized audio and reduce spectral defects. Thus, the hidden representation \textit{r} of the signal can be defined as:
\begin{align}
    r = \sigma({f}_{1}(sp)) \otimes emb(pitch) \otimes sp + {f}_{2}(sp)
\end{align}
Here, ${f}_{1}$ denotes the ResBlock1 and ${f}_{2}$ denotes the ResBlock2. \textit{sp} denotes the spectral envelope in full linear scale and \textit{emb} denotes the embedding representation of input pitch. In the ResBlocks, we set the dilation rates to [1, 2, 1, 2], and the kernel sizes to 3.

%%%%%%%%%%%%%%%%%%%%%%%%%%%%%%%%%%%%%%%%%%%%%%%%%%%%%%%%%%%%%
\subsubsection{Fre-GAN Vocoder}
Fre-GAN \cite{kim2021fregan} is a neural network vocoder with feed-forward transposed convolution blocks up-sampling the input mel-spectrogram until the output reaches the expected waveform sampling rate. It outperforms auto-regressive neural vocoders in inference speed, unseen-speaker generalization, and pitch consistency, which meets the requirements for the pitch correction system.
%while preserve high-frequency resolution. 

% The main contribution of Fre-GAN is that it replaces the average-pooling with Discrete Wavelet Transform (DWT). Multi-Scale Discriminator in previous works \cite{kong2020hifi}\cite{kumar2019melgan} implemented the down-sample function as strided average pooling, which will blur the high frequency component and introduce distortion in high frequency bands in the generated waveform. To provide a solution to this problem, Fre-GAN replaces average pooling with the DWT as the new down-sampling method in RSD. The waveform go through the low-pass filter and the high-pass filter respectively and the outputs are concatenated channel-wise to perform down-sampling without information loss.

In the generator, a multi-receptive field fusion (MRF) module proposed in HiFi-GAN \cite{kong2020hifi} is employed to observe patterns on diverse scales. Skip-connections and up-sampling modules are also adopted at top-K deep layers to sum up different sample rates' features to increase resolution gradually and stabilize the adversarial training process. 
The overall architecture is called the Resolution-Connected Generator (RCG) block. 
In our work, the input of the RCG block is the hidden representation from SF block rather than the mel-spectrogram. Since the sampling rate of our experiment is different from the original Fre-GAN, we also modified some of the parameters in the up-sampling layers.

Two discriminators from the Fre-GAN are also employed in KareTuner, including the Resolution-wise multi-Period Discriminator (RPD) and Resolution-wise multi-Scale Discriminator (RSD)\footnote{We used the implementation of the discriminators in: \url{https://github.com/rishikksh20/Fre-GAN-pytorch}, although it is not exactly the same as the original paper.}. %Fre-GAN achieves higher sound quality than many other neural vocoders, because it replaces the average-pooling with Discrete Wavelet Transform (DWT). 
There, Discrete Wavelet Transform (DWT) instead of average pooling is applied to the waveform to achieve down-sampling without information loss.
%%%%%%%%%%%%%%%%%%%%%%%%%%%%%%%%%%%%%%%%%%%%%%%%%%%%%%%%%%%%%
\begin{table*}[h]
\centering
\caption{MOS evaluation results with their 95\% confidence intervals and the root-mean-square of the pitch error in cents}
\begin{tabular}{lccc}
\toprule
\textbf{Model}   & \textbf{sound quality MOS} $\uparrow$  & \textbf{overall MOS} $\uparrow$  & \textbf{F0 RMSE} $\downarrow$ \\
\midrule
Phase Vocoder      & $3.07\pm0.25$     & $3.01\pm0.22$      & $19.0$\\
WORLD              & $3.69\pm0.26$     & $3.79\pm0.17$      & $\bm{17.2}$\\
CLPCNet            & $3.80\pm0.22$     & $3.81\pm0.21$      & $69.0$\\
\midrule
\textbf{Ours}      & $\bm{4.19\pm0.22}$ & $\bm{4.19\pm0.15}$   & $38.3$ \\
\bottomrule
\end{tabular}
\label{table:MOS}
\end{table*}
%%%%%%%%%%%%%%%%%%%%%%%%%%%%%%%%%%%%%%%%%%%%%%%%%%%%%%%%%%%%%
\subsubsection{Training Objectives}
The training of the activation model and the vocoder was conducted in an end-to-end manner, and the network is optimized to reconstruct the real waveform from ground-truth pitch curve and spectral envelope.

The generator loss is defined as:
\begin{align}
\begin{split}
    \mathcal{L}_G &= \sum_{n=0}^4\mathbb{E}[||D_n^P(\hat{x}) - 1||_2 + \lambda_{fm}\mathcal{L}_{fm}(G;D_n^P)] \\
    &+ \sum_{n=0}^2\mathbb{E}[||D_n^S(\hat{x}) - 1||_2 + \lambda_{fm}\mathcal{L}_{fm}(G;D_n^S)] \\
    &+ \lambda_{STFT}\mathcal{L}_{STFT}(G)
\end{split}
\end{align}

The discriminator loss is defined as:
\begin{align}
\begin{split}
    \mathcal{L}_D &= \sum_{n=0}^4\mathbb{E}[||D_n^P(x) - 1||_2 + ||D_n^P(\hat{x})||_2] \\
    &+ \sum_{n=0}^2\mathbb{E}[||D_m^S(\phi^m(x) - 1)||_2 + ||D_m^S(\phi^m(\hat{x}))||_2]
\end{split}
\end{align}
Here, $x$ denotes the ground truth waveform, $\hat{x}$ denotes the generated waveform, $G$ denotes the SF layer and RCG, $D_n^P$ denotes the $n$-th RPD, $D_n^S$ denotes the $n$-th RSD, $\phi^m$ denotes the $m$-th level DWT, $\lambda_{fm}$ and $\lambda_{STFT}$ are weighting parameter for feature loss $\mathcal{L}_{fm}$ and STFT-spectrogram loss $\mathcal{L}_{STFT}$ respectively. The lambda parameters aim to balance the generative and adversarial losses in different scales. According to our experiments, these parameters are not particularly strict, but improper parameter settings usually make the training process unstable and introduce artifacts in the generated results. In the experiments, we set $\lambda_{fm}$ = 2
and $\lambda_{{STFT}}$ = 45 which balance the adversarial losses.

The feature loss is defined as:
\begin{align}
    \mathcal{L}_{fm}(G; D_k) = \mathbb{E}\left[\sum_{i=0}^{T-1}\frac{1}{N_i}||D_k^{(i)}(x) - D_k^{(i)}(\hat{x})||_1\right]
\end{align}

Where $D_k^{(i)}$ denotes the $i$-th feature extracted by discriminator $D_k$.

The STFT-spectrogram loss is defined as:
\begin{align}
    \mathcal{L}_{STFT}(G) = \mathbb{E}[||\psi(x) - \psi(\hat{x})||_1]
\end{align}

Where $\psi(x)$ denotes the STFT function.
%%%%%%%%%%%%%%%%%%%%%%%%%%%%%%%%%%%%%%%%%%%%%%%%%%%%%%%%%%%%%
\section{Experiments}
\subsection{Dataset and Data Preprocessing}

In the pitch correction task, there hardly exists paired data that includes both out-of-tune and in-tune vocals of a song from the same singer, which increases the difficulty in training. Therefore, this paper's novelty is that we conducted HMM smoothing \cite{matthias2014a} \cite{m2015a} to the out-of-tune vocals to extract standard MIDI note sequence as the reference note template in the training data. In the training phase, our model learns to generate the out-of-tune pitch curve from the corresponding out-of-tune notes. In the inferencing phase, we replace the note sequence with the target musical notes which will lead to in-tune pitch outputs. In this method, we built a large dataset without manual labeling to complete the pitch prediction task.

We collected 5294 full-song performances by amateur singers of different singing proficiency in karaoke settings, which are time-aligned with the accompaniment, with an average of 4.3 minutes. The same dataset is also used in vocoder training.
%%%%%%%%%%%%%%%%%%%%%%%%%%%%%%%%%%%%%%%%%%%%%%%%%%%%%%%%%%%%%
\subsection{Experiments Settings}
\label{sec:Experiments Settings}
The spectral envelopes are extracted with cheaptrick algorithm in WORLD vocoder \cite{Morise2016WORLDAV} with 2048 of window size, 512 of hopsize, and 2048 points of Fourier transform.

To meet the sound quality requirement of music production, we raised the sampling rate of the synthesized waveform from 22050Hz to 32000Hz, and STFT hopsize from 256 to 512. The up-sampling rate of the transposed convolution layers are set to [8, 4, 4, 2, 2], the kernel sizes are set to [16, 8, 8, 4, 4], and dilation rates of MRF are set to [[1, 1], [3, 1], [5, 1], [7, 1] × 3]. We used AdamW optimizer with $\beta_1=0.8$, $\beta_2=0.99$, $batch\_size=128$.

To evaluate the performance of the proposed method and the baseline, we ask 12 people with good music training experience to do the subjective test. We used 13 audio clips with lengths from 5s to 10s, and each candidate was randomly assigned four clips to evaluate pitch predictor performance and other four clips to evaluate vocoder performance\footnote{The audio examples are available at: 

\url{https://ella-granger.github.io/KaraTuner}}.
%%%%%%%%%%%%%%%%%%%%%%%%%%%%%%%%%%%%%%%%%%%%%%%%%%%%%%%%%%%%%
\subsection{Experiment 1: Pitch Predictor Performance}
\label{ssec:pitch predictor performance}
\begin{figure}[ht]
  \centerline{\includegraphics[width=7cm]{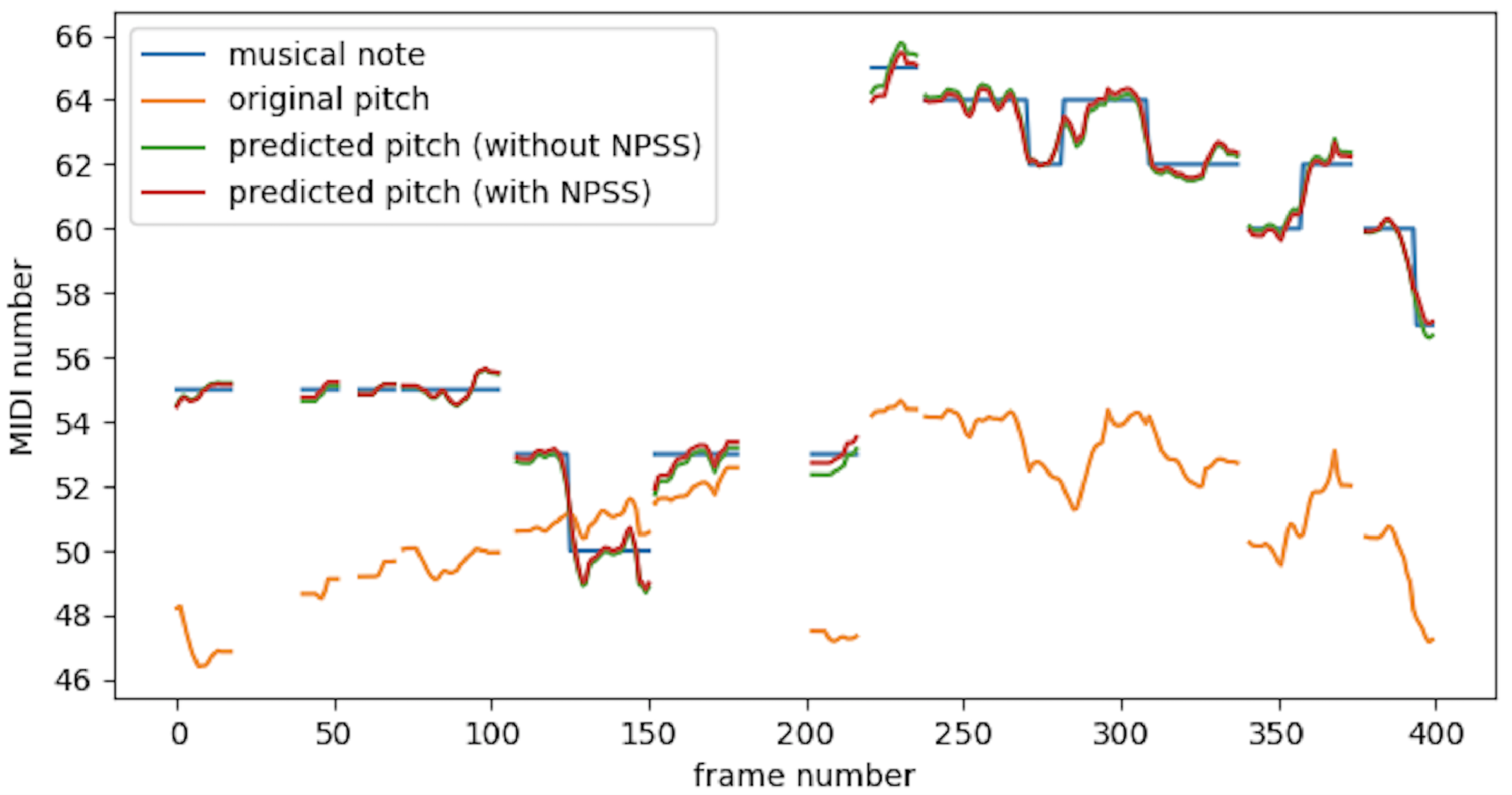}}
  \caption{An example from the results of the proposed pitch predictor.}
\label{fig:pitch}
\end{figure}

We used the post-tuning process in NPSS\cite{blaauw2017neural} as the pitch tuning baseline, which is a note shifting algorithm. It iterates through every note in the reference score, and moves the corresponding pitch curve to eliminate the difference between the estimated average of the curve and the target note. In this way, it performs pitch correction without altering the details such as bending and vibratos in the original curve. This method was also applied to the predicted pitch curve to obtain perfect intonation. Figure \ref{fig:pitch} illustrates an example of the musical note, the original pitch curve, the predicted pitch curve with and without NPSS post-tuning. Here, the original pitch means the pitch curve extracted from the vocals by karaoke singers, which we can assume that they are usually out of tune. The predicted pitch means the pitch curve estimated from KaraTuner, which we hope they are in-tune and match the input musical notes. Audios in this test were all synthesized with our proposed vocoder.

\begin{figure}[ht]
  \centerline{\includegraphics[width=7cm]{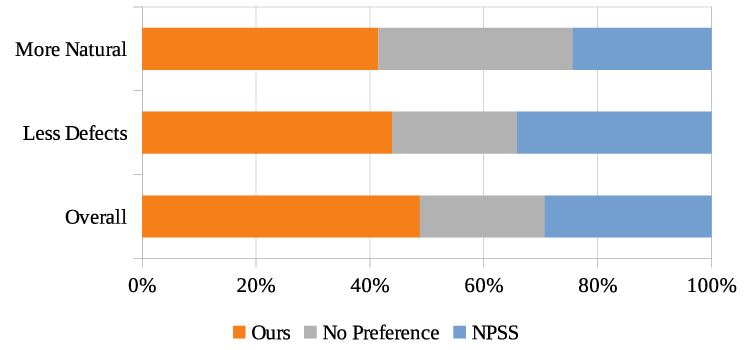}}
  \caption{Experiment 1: A/B test on pitch predictor.}
\label{fig:ABtest}
\end{figure}

We conducted A/B tests on pitch naturalness, the number of defects, and overall performance between the proposed pitch predictor and the baseline method. We collected 41 valid answers, and the results in Figure \ref{fig:ABtest} show that the raters prefer our proposed method in all three criteria.

Since both curves went through the post-tuning method in \cite{blaauw2017neural}, the differences in user preference lie in the details of the pitch curves. We observe that the predictor removes imperfect slides and shakes in the original pitch curve, while generating smoother transitions between notes.

\subsection{Experiment 2: Vocoder Performance}
%%%%%%%%%%%%%%%%%%%%%%%%%%%%%%%%%%%%%%%%%%%%%%%%%%%%%%%%%%%%%（这一段帮忙看看）

We used the phase vocoder, WORLD vocoder and CLPCNet as baselines to synthesize the pitch-corrected audio. Mean Opinion Scale (MOS) evaluations were conducted over sound quality and the overall quality considering the timbre consistency, and the results of 43 valid answers are shown in Table \ref{table:MOS}. In the subjective evaluation, the proposed vocoder achieved the highest MOS score in both sound quality and the overall quality, which proves the significant advantage of the source filter block and the neural vocoder. In our objective evaluation of pitch accuracy, we find that traditional DSP vocoders have significant advantage than neural network vocoders, but our proposed vocoder has a lower root-mean-square of the pitch error than the CLPCNet.
%%%%%%%%%%%%%%%%%%%%%%%%%%%%%%%%%%%%%%%%%%%%%%%%%%%%%%%%%%%%%
\section{CONCLUSION}
In this paper, we proposed KaraTuner which performs end-to-end pitch correction. It predicts a natural pitch curve from the spectral envelope and a score reference, then synthesizes high fidelity in-tune singing voice while maintaining the original audio's timbre. Experiment results suggest that evaluators show a stronger preference for KaraTuner than other baseline solutions. For future work, we will continue to optimize the quality in scenes of reverberation, noise and inaccurate rhythm of singing vocal.

\vfill\pagebreak

\bibliographystyle{IEEEtran}

\bibliography{mybib}

\end{document}